# A Langmuir-Like Adsorption Model for $MoS_2$ Gas Sensors: Bridging Theory and Experiment

Z. Geng[1], M. Ziegler[2], and F. Schwierz[1]
[1]Department of Micro- and Nanoelectronic Systems, Technische Universität Ilmenau, Germany
[2]Department of Materials Science, Kiel University, Germany

**Abstract**

Gas sensors based on layered TMDCs (transition-metal dichalcogenides) such as $MoS_2$ have attracted considerable attention recently and are considered as a promising alternative to the conventional metal-oxide-based gas sensing devices. Recent studies on TMDC gas sensors have mainly addressed either the theoretical description of adsorption mechanisms or the fabrication of test devices and the investigation of their gas sensing performance. However, a suitable approach for modeling the gas sensing behavior of TMDC devices and for establishing a reliable connection between theoretical results and the response of experimental devices when exposed to gases remains absent. To bridge this gap, in the present work a new Langmuir-based adsorption model, capable of calculating both the sensor sensitivity and selectivity, is developed and applied to $MoS_2$-based gas sensors. It is shown that by using appropriate published data for the adsorption energy and the charge transfer as input data, the reported experimental sensitivities of $MoS_2$ gas sensors can be well reproduced by the model.

## 1. Introduction

Increasing industrial activities and vehicle exhaust emissions have led to elevated atmospheric concentrations of harmful and toxic gases such as CO, $CO_2$, NO, $NO_2$ and $NH_3$ [1]. For instance, annual mean $NO_2$ concentrations in some urban areas exceed the WHO guideline value of 10 µg $m^{-3}$ (≈ 5.3 ppb) [2]. These gases pose serious risks to human health and contribute to air pollution, acid rain, and the greenhouse effect. Consequently, the development of reliable gas sensors capable of detecting and monitoring trace-level gaseous pollutants has become increasingly important [1,3–6].

| Type of gas sensor | Sensitivity (e.g. $NO_2$) | LOD for $NO_2$ | Selectivity | Energy consumption |
|---|---|---|---|---|
| Thermoelectric | Low [7–10] | 2 ppm [9] | Moderate [7–10] | High/Low [7–10] |
| Mass-sensitive | Moderate [11–13] | 175 ppb [13] | Moderate [11–13] | Moderate [11–13] |
| Magnetic | Moderate [14–16] | 150 ppb [16] | Moderate [14–16] | Moderate [14–16] |
| Electrochemical | High [17–19] | 4.8 ppb [19] | Moderate [17–19] | Low [17–19] |
| Optical | High [20–22] | 1 ppb [22] | High [20–22] | High [20–22] |
| Electronic | High [23–26] | 0.05 ppb [26] | Moderate [23–26] | Low [23–26] |

Table 1. Comparison of thermoelectric, mass-sensitive, magnetic, electrochemical, optical, and electronic gas sensors in terms of sensitivity, LOD for $NO_2$, selectivity, and energy consumption. The listed LODs for $NO_2$ are experimentally reported unless otherwise stated. For electronic gas sensors, the lowest LOD is theoretically estimated. The qualitative classifications of sensitivity, selectivity, and energy consumption reflect typical performance trends reported in the literature [7–26].

Traditional gas sensors can be categorized into six types based on their operating principles: thermoelectric, mass-sensitive, magnetic, electrochemical, optical, and electronic [4,6]. Among these categories, electronic gas sensors offer several advantages, including high sensitivity, reasonable selectivity, low power consumption (typically below a few milliwatts), relatively simple fabrication and operation, low cost, compact size, compatibility with IoT (Internet of Things) systems, and long-term stability [4,23–28]. Table 1 presents a comparison of different gas sensor types, mainly focusing on sensitivity, LOD (limit of detection) for $NO_2$ concentration, selectivity, and energy consumption [17–26]. In terms of sensitivity, using $NO_2$ as an example, thermoelectric gas sensors demonstrate the lowest detection capability, with a LOD of 2 ppm [9]. Mass-sensitive and magnetic gas sensors can detect $NO_2$ at concentrations of 175 ppb [13] and 150 ppb [16], respectively. Electrochemical and optical sensors follow, achieving LODs of 4.8 ppb and 1 ppb [19,22], respectively. In contrast, electronic gas sensors exhibit the highest detection capability, with a theoretically estimated LOD reaching as low as 0.05 ppb [26].

In conventional electronic gas sensors, the sensing layers usually consist of semiconducting metal oxides such as tin, tungsten, zinc, indium, or titanium oxide, and have, depending on the deposition method, typical thicknesses ranging from several nm up to the µm range. For a high sensitivity and a small LOD, the surface-to-volume ratio of the sensing layer should be large, thus thin sensing layers are preferable.

2D (two-dimensional) materials such as semiconducting mono- or few-layer TMDCs (transition metal dichalcogenides) offer unmatched surface-to-volume ratios since they are ultimately thin. Monolayer $MoS_2$, for example, has a thickness of only 0.65 nm. Therefore, 2D TMDCs are considered as promising candidates for the sensing layer of future high-sensitivity electronic gas sensors [5,29–31].

Numerous experimental TMDC-based gas sensors, particularly those employing $MoS_2$, have been reported and their response to gases has been investigated [32–44]. On the theoretical side, extensive efforts have been devoted to *ab initio*, i.e., first-principles, simulations using DFT (density functional theory) to investigate the fundamental principles (or mechanisms) of the adsorption of gas molecules on TMDC surfaces and to calculate characteristic adsorption parameters such as the adsorption energy $E_{ads}$ and the resulting charge transfer $\Delta Q$ [45–58]. On the other hand, limited attention has been paid to activities related to modeling gas adsorption on 2D materials at the device level and to describing the response of 2D gas sensors to the gas exposure.

In the present work, a modeling approach to describe the electrical response of $MoS_2$-based gas sensors to ambient gas exposure is proposed. It is based on the Langmuir adsorption isotherm, combines essential theoretical concepts of gas adsorption on $MoS_2$ surfaces, and subsequently relates the modeled sensor response to that of experimental sensors reported in the literature. The proposed model effectively and quantitatively bridges the gap between the first-principles investigations of gas adsorption and experimental results covering low and high gas concentration ranges. Validation using $NH_3$ and $NO_2$ as representative donor- and acceptor-like gases demonstrates excellent agreement with experimental data, indicating the reliability and predictive capabilities of the proposed model. The methodology can easily be extended to TMDC gas sensors with sensing layers made of semiconducting TMDCs beyond $MoS_2$. We note, however, that our approach also faces inherent problems. In particular, key input parameters required by the model, such as the adsorption energy $E_{ads}$ and the charge transfer $\Delta Q$, adopted from the literature may vary significantly depending on the theoretical methods and the material models used in different studies. This variability introduces an uncertainty into the predicted sensor response.

## 2. Langmuir-Based Model for the Gas Adsorption

### 2.1 Device Structure and Operating Principle

A schematic illustration of a gas sensor using monolayer $MoS_2$ as the sensing layer is shown in Fig. 1. The device consists of a defect-free $MoS_2$ channel with ideal Ohmic Au contacts on a thermally oxidized Si wafer. By analogy with the field-effect transistor, we call these contacts source and drain. A metal electrode located at the bottom of the Si substrate serves as a back gate. In the absence of gas adsorption, as shown in Fig. 1 (a), the $MoS_2$ channel remains in a pristine state, and the flowing drain current $I_{D\text{-}0}$ is determined by the intrinsic carrier concentration, the applied drain-source voltage $V_{DS}$, and the electrostatic back-gate modulation.

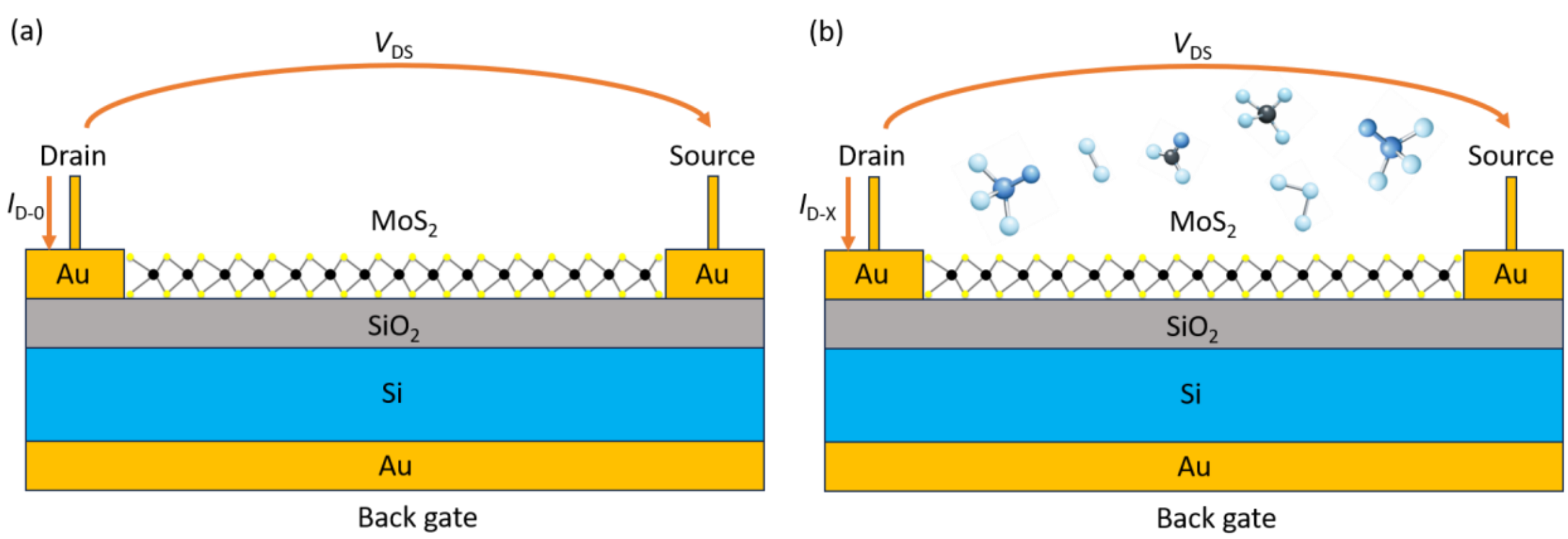

Figure 1. Schematic of a $MoS_2$-based gas sensor, after [32], (a) in the absence of gas adsorption and (b) upon gas adsorption onto the $MoS_2$ surface.

When exposed to gas molecules, as illustrated in Fig. 1 (b), gas molecules are adsorbed on the $MoS_2$ surface, leading to a charge transfer between the adsorbed gas molecules and the channel. This adsorption-induced interaction modifies the carrier concentration and conductivity of the channel, resulting in a change of the drain current from $I_{D\text{-}0}$ to $I_{D\text{-}X}$, at the same applied drain-source voltage $V_{DS}$. This current variation reflects the adsorption-induced changes occurring on the TMDC surface.

### 2.2 Gas Adsorption Modeling

As indicated above, the sensing operation of $MoS_2$ gas sensors, as that of conventional metal-oxide-based electronic gas sensors, is attributed to adsorption-induced changes in the conductivity of the channel. For fixed bias conditions, the sensor performance can be characterized by analyzing the difference between the behavior of the sensor without the influence of gases (see Fig. 1(a)) and under the influence of gases as shown in Fig. 1(b).

The first step of the model development is to express the surface density of adsorbed gas molecules $\rho_{ads\text{-}H}$ using the simple Henry adsorption isotherm under the assumption that the $MoS_2$ surface is in thermodynamic equilibrium with the surrounding gas. Thus, transient effects are not taken into account. Pu et al. [46] proposed the expression

$$\rho_{ads\text{-}H} \approx \frac{5}{2}\frac{P_X \lambda_{th\text{-}X}}{k_B T}\left[\exp\left(\frac{-E_{ads\text{-}X}}{k_B T}\right)\right]^2 = P_X \cdot k_H \tag{1}$$

for the calculation of $\rho_{ads\text{-}H}$, where $P_X$ represents the partial pressure of the gas of species X, $E_{ads\text{-}X}$ is the adsorption energy of an isolated gas molecule, $k_B$ is the Boltzmann constant, $T$ is the temperature, and $\lambda_{th\text{-}X}$ is the thermal de Broglie wavelength given by

$$\lambda_{\text{th-X}} = \frac{h}{\sqrt{2\pi m_X k_B T}} \tag{2}$$

with $h$ being Planck's constant and $m_x$ the mass of a single molecule of the gas of species X. The Henry isotherm assumes a linear dependence of $\rho_{\text{ads-H}}$ on $P_X$, implying that the number of adsorbed gas molecules continuously increases with rising gas concentration. Such a linear relationship describes the reality well for low gas concentrations but neglects the fact that the number of surface states available for adsorption is limited and that the density of adsorbed gas molecules saturates at high gas concentrations. To take this saturation into account, another adsorption isotherm has to be used. Widely used for this purpose is the Langmuir adsorption isotherm [47].

Figure 2 shows a qualitative comparison of the adsorption characteristics according to the Henry and the Langmuir isotherms. As can be seen, both adsorption curves coincide at low gas concentrations but diverge more and more at higher concentrations.

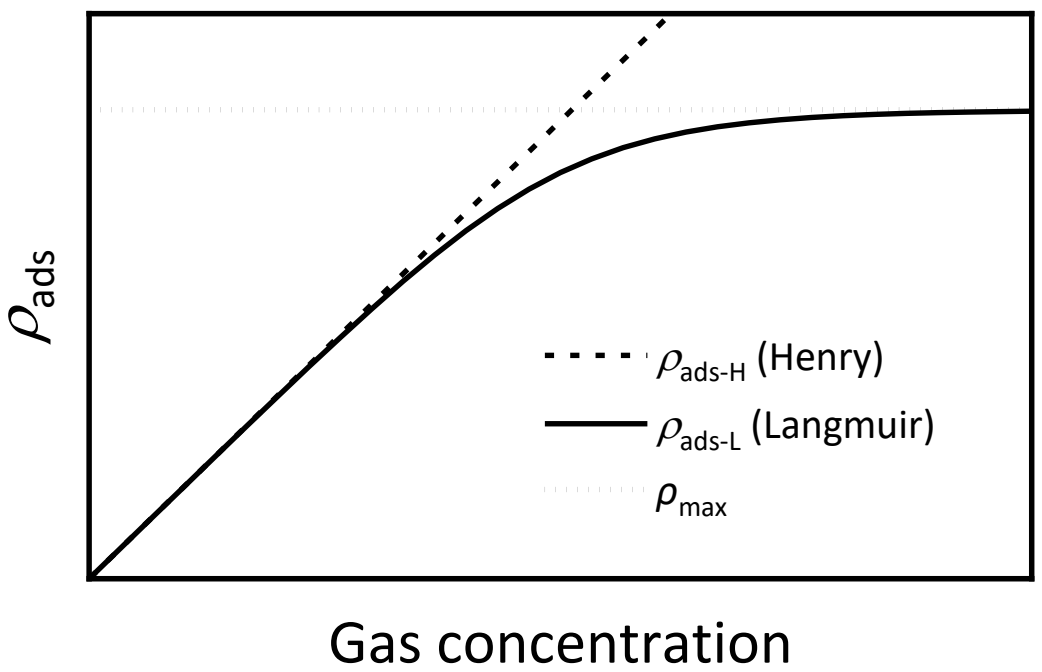

Figure 2. Qualitative comparison of the density of adsorbed gas molecules according to the Henry ($\rho_{\text{ads-H}}$) and Langmuir ($\rho_{\text{ads-L}}$) isotherms.

The dependence of the density of adsorbed gas molecules $\rho_{\text{ads-L}}$ on the gas concentration shown in Fig. 2 has the same general shape as the dependence of the charge-carrier drift velocity in Si on the electric field, which can be described by the empirical Caughey-Thomas formula [48]. Motivated by this analogy, we express the density of adsorbed gas molecules by a modified Caughey-Thomas formula as

$$\rho_{\text{ads-L}} = \rho_{\text{ads-H}} \left( \frac{1}{1+\left[\frac{\rho_{\text{ads-H}}}{\rho_{\text{max}}}\right]^{\beta}} \right)^{\frac{1}{\beta}} \tag{3}$$

where $\rho_{\text{ads-H}}$ and $\rho_{\text{ads-L}}$ are the density of adsorbed gas molecules obtained from the Henry isotherm according to eq. (1) and the density obtained from our Langmuir-like model, $\beta$ is a bowing parameter determining how quickly (in terms of gas concentration) the curve goes into saturation, and $\rho_{\text{max}}$ represents the maximum adsorption density. In the present work, $\beta$ is set to 0.5 and for the calculation of $\rho_{\text{max}}$ we assume that every unit cell of the monolayer $MoS_2$ sensing layer provides one single adsorption site. Considering the hexagonal $MoS_2$ lattice with a lattice constant $a$ = 3.16 Å (1 Å = $10^{-10}$ m) [49], the maximum adsorption density is given by

$$\rho_{\text{max}} = \frac{1}{A_{\text{unitcell}}} = \frac{2}{\sqrt{3}\times(3.16\times10^{-10})\times10^{4}}\text{cm}^{-2} = 1.16\times10^{15}\text{cm}^{-2} \tag{4}$$

where $A_{\text{u-cell}}$ is the area of the unit cell of the $MoS_2$ lattice.

### 2.3 Two Key Parameters for Adsorption Modeling

Having described the main features of the new Langmuir-like adsorption model, next we discuss two key parameters that are essential for modeling gas sensor performance. The first one is the adsorption energy $E_{\text{ads-X}}$ of the considered gas species X. Since gas adsorption is an exothermic process, the adsorption energy is negative and a more negative $E_{\text{ads}}$ indicates a more favorable adsorption process, i.e., stronger (or more) adsorption. Furthermore, the charge transfer $\Delta Q_X$ of the gas under consideration must be known. Gas molecules adsorbed to a semiconductor surface behave either donor-like or acceptor-like. A donor-like gas transfers negative charge to the semiconductor, thereby increasing the conductivity of an n-type semiconductor or decreasing the conductivity of a p-type semiconductor. Such a behavior is indicated by positive charge transfer. Acceptor-like gases, on the other hand, behave in the opposite way. They withdraw negative charge from the semiconductor and are characterized by a negative $\Delta Q$.

Reported data for $E_{\text{ads}}$ and $\Delta Q$ for gas adsorption on $MoS_2$ are usually the results of density functional theory simulations [50–58] while experimental data are difficult to obtain. Figure 3 shows the adsorption energies for various gases on $MoS_2$ compiled from the literature and reveals a wide variation of $E_{\text{ads}}$ for identical gas species. As can be seen, the reported $E_{\text{ads}}$ for $NH_3$ range from −0.16 eV to −0.4 eV, while for $NO_2$ the reported $E_{\text{ads}}$ cover a range from −0.14 eV to −0.27 eV [34,50–55]. Such differences in $E_{\text{ads}}$ lead to significant variations in the modeled sensor performance. Therefore, before modeling a $MoS_2$ gas sensor, the primary task is to select reasonable $E_{\text{ads}}$ data for the gases under consideration.

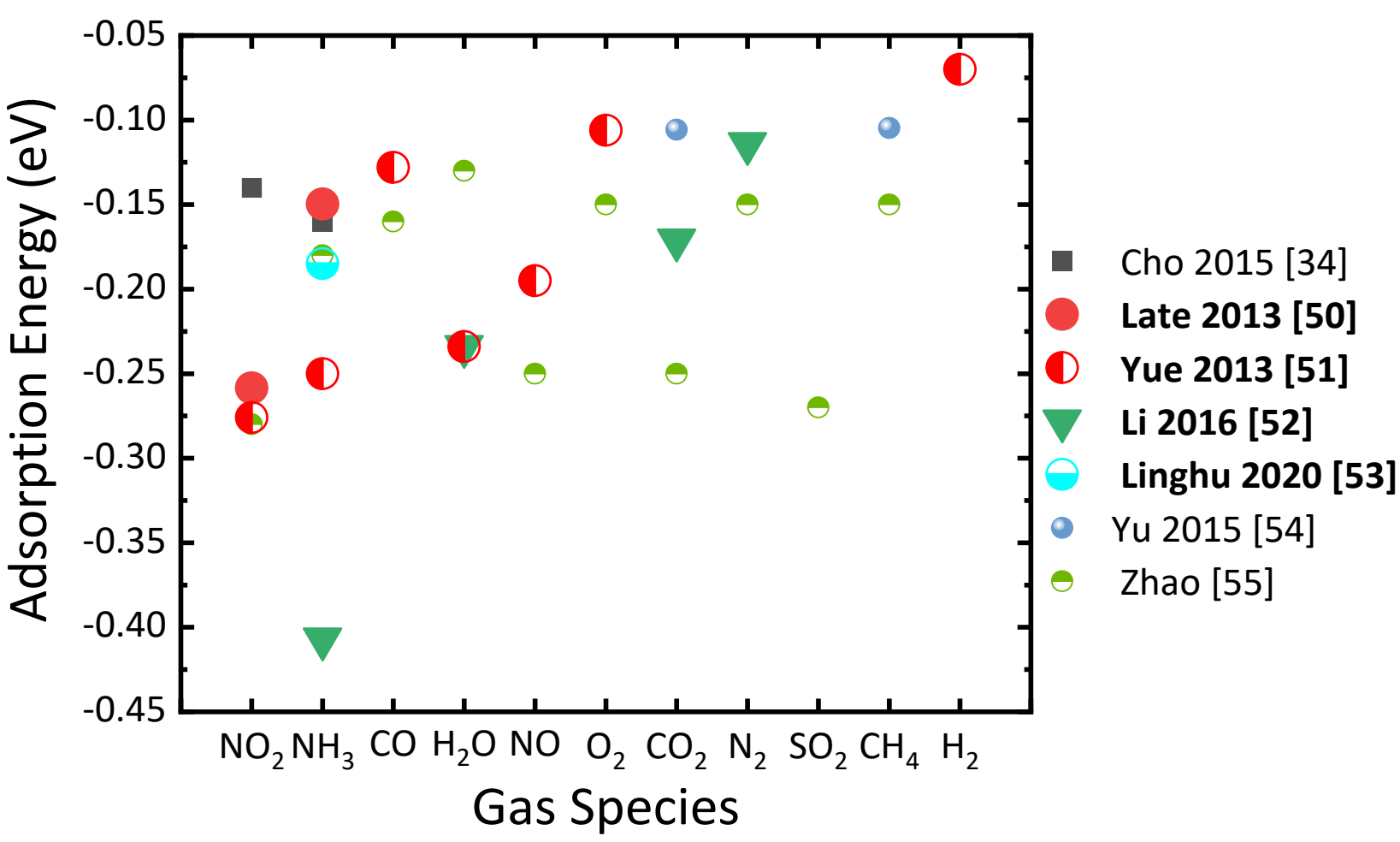

Figure 3. Adsorption energies of various gases on $MoS_2$ reported in the literature [34,50–55]. The values highlighted by large symbols are taken from four representative studies that report both $E_{\text{ads}}$ and the corresponding charge transfer $\Delta Q$ for $NH_3$, enabling a consistent parameter set for the subsequent modeling analysis.

Figure 4 shows the compiled charge-transfer data between different gas molecules and the $MoS_2$ surface. As with $E_{\text{ads}}$, the reported $\Delta Q$ data for identical gas species also vary widely. The effect of $\Delta Q$ on the electronic properties of the $MoS_2$ channel is quantitatively described by eq. (5), which defines the change in carrier density $\Delta n$ induced by adsorption [51].

$$\Delta n = \frac{\rho_{\text{ads-L}} \cdot \Delta Q}{t} \tag{5}$$

Here, $\Delta Q$ denotes the charge transfer per gas molecule and $t$ represents the thickness of the $MoS_2$ channel.

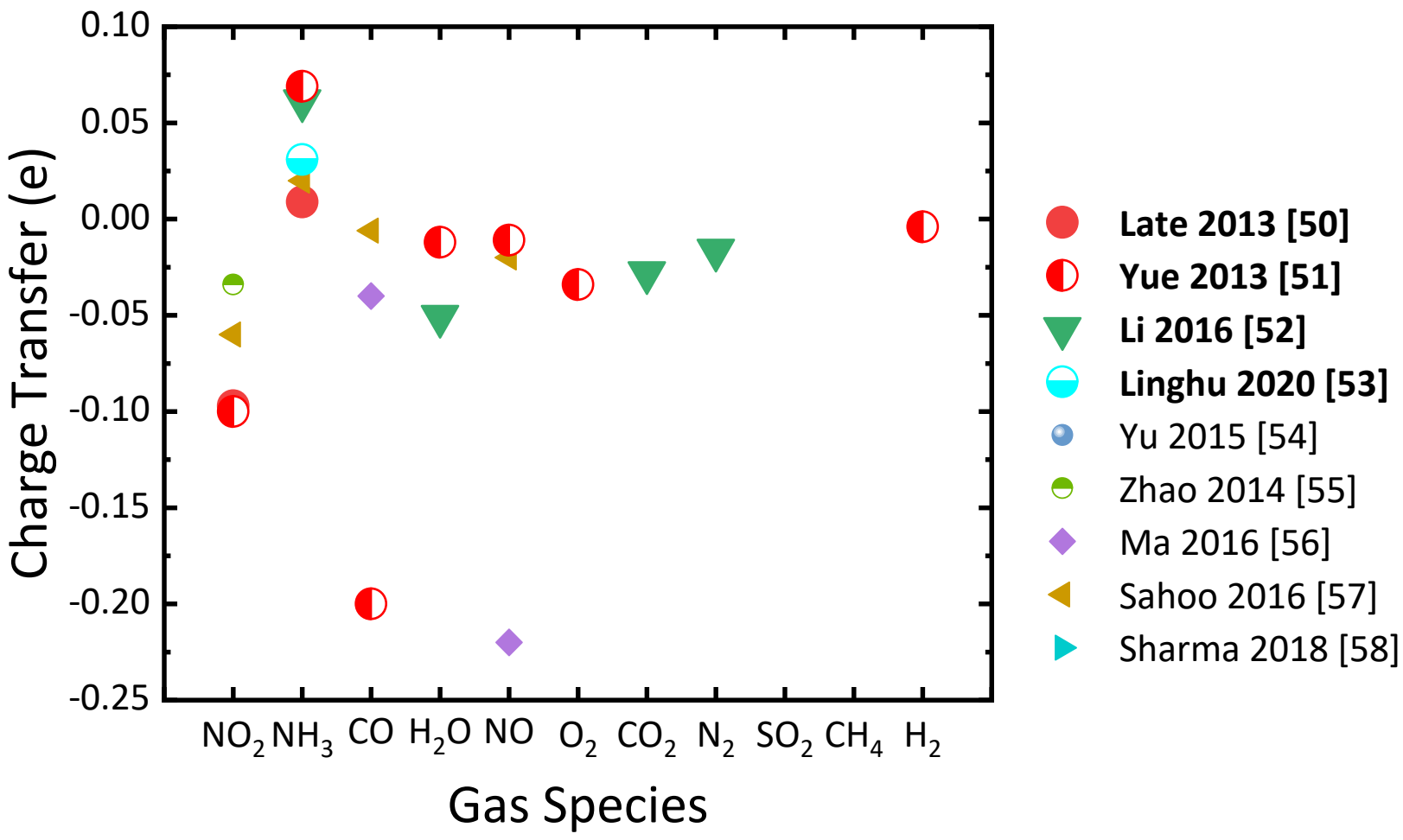

Figure 4. Charge transfer data in units of the elementary charge $q$ for the adsorption of different gases on $MoS_2$ compiled from the literature [50–58]. The $\Delta Q$ data highlighted by large symbols were selected for our modeling analysis since in the corresponding papers both $\Delta Q$ and $E_{ads}$ are reported.

Pristine $MoS_2$ is generally an n-type semiconductor due to intrinsic defects like sulfur vacancies acting as donors. Consequently, gases with $\Delta Q > 0$, such as $NH_3$, increase the electron concentration and conductivity, whereas gases with $\Delta Q < 0$, such as $NO_2$, extract electrons and reduce the conductivity.

Throughout much of the present work, $NH_3$ and $NO_2$ are selected as representative donor- and acceptor-type gases to illustrate the proposed modeling approach. Other gas species are discussed later in the paper in the frame of a selectivity analysis. Equation (5) is used to determine the change in carrier density $\Delta n$.

## 3. Sensor Modeling and Simulation

As discussed above, the adsorption of gas molecules on the $MoS_2$ surface induces a variation of the carrier concentration in the channel, which is reflected by a corresponding change in the drain current $\Delta I$. The sensitivity $S_G$ of a gas sensor is defined by the corresponding relative change of the sensor conductance $\Delta G$ between the case where a gas of species x is adsorbed on the channel surface and the case without gas adsorption. Provided that the same voltage $V_{DS}$ is applied to the sensor in the two cases, the sensitivity is expressed by

$$S_G = \left|\frac{\Delta G}{G_{D\text{-}0}}\right| = \left|\frac{G_{D\text{-}X} - G_{D\text{-}0}}{G_{D\text{-}0}}\right| = \left|\frac{\Delta I}{I_{D\text{-}0}}\right| = \left|\frac{I_{D\text{-}X} - I_{D\text{-}0}}{I_{D\text{-}0}}\right| = \left|\frac{I_{D\text{-}X}}{I_{D\text{-}0}} - 1\right| \tag{6}$$

where the indices D-0 and D-X indicate the cases without and with gas adsorption, respectively. In the present work, the currents $I_{D\text{-}0}$ and $I_{D\text{-}X}$ are calculated by two complementary approaches, namely

(1) by analytical modeling and
(2) by numerical device simulation using a commercial TCAD tool,

always at room temperature.

In the frame of approach (1), the drain current without gas adsorption is modeled by

$$I_{\text{D-0}} = \frac{q \times n_0 \times \mu_0 \times V_{\text{DS}} \times W \times t}{L} \tag{7}$$

where $q$ is the elementary charge, $n_0$ denotes the initial carrier concentration equal to the channel doping, $\mu_0$ is the carrier low-field mobility, and $L$, $W$, and $t$ represent the length, width, and thickness of the $MoS_2$ channel, respectively. Thus, the saturation of the carrier drift velocity at high electric fields is neglected, a simplification justified by the long channels of experimental $MoS_2$ gas sensors and by the rather moderate carrier mobility in $MoS_2$. The drain current with gas adsorption, on the other hand, is calculated using

$$I_{\text{D-X}} = \frac{q \times (n_0 + \Delta n) \times \mu_0 \times V_{\text{DS}} \times W \times t}{L} \tag{8}$$

where $\Delta n$ is the variation of the channel carrier concentration caused by the charge transfer between the adsorbed gas molecules and the channel. Finally, the sensitivity of the sensor is obtained from

$$S_{\text{G}} = \left|\frac{\Delta G}{G_{\text{D-0}}}\right| = \left|\frac{\Delta I}{I_{\text{D-0}}}\right| = \left|\frac{\Delta n}{n_0}\right| \tag{9}$$

The second approach involves device-level simulation using the ATLAS module of the Silvaco TCAD environment, which is designed for semiconductor device analysis [59]. Since TMDCs are not included in the default ATLAS material database, the $MoS_2$ material parameters needed for the simulations must be explicitly defined. These parameters are compiled in Tab. 2.

|  |  |  |  |
|---|---|---|---|
| Bandgap [eV] | 1.8 [60] | Saturation velocity $v_{sat}$ [cm s$^{-1}$] | $1 \times 10^7$ [62] |
| Mobility $\mu$ [cm$^2$ V$^{-1}$ s$^{-1}$] | 125 [61,62] | Fitting parameter $\beta_v$ for $v_{sat}$ | 1.2 [61] |
| Dielectric constant $\varepsilon_r$ | 3.93 [60] | Correction factor | 4 [61] |
| Effective mass (electron) | 0.463$m_0$ [60] | Effective mass (Hole) | 0.557$m_0$ [60] |

Table 2. $MoS_2$ material parameters applied in the present work for sensor modeling and simulation. Note that (i) For analytical modeling only the low-field mobility from the table is used; (ii) The parameter $\beta_v$ is needed for the carrier transport model in ATLAS; (iii) To properly simulate the density of states in the 2D $MoS_2$ channel, the actual electron density of states effective mass of bulk $MoS_2$ is modified by the correction factor $C$; (iv) $m_0$ is the electron rest mass.

The effect of gas adsorption is described in both approaches (1) and (2) through the adsorption-induced change of the carrier concentration in the channel. Specifically, the charge transfer between adsorbed gas molecules and the $MoS_2$ channel gives rise to a carrier density variation $\Delta n$. In the analytical modeling, this change is directly used to evaluate the sensitivity via eq. (9). In the numerical simulations, the effect of the charge transfer is implemented by a uniform modification of the effective doping concentration in the $MoS_2$ channel, thereby ensuring consistency between the analytical and numerical descriptions of gas-induced current modulations.

## 4. Results

### 4.1 Comparisons of Modeling – Simulation – Experiment

To assess whether our theoretical approach to describe the effects of gas adsorption is able to reproduce the behavior of real gas sensors, a comparison between theory and experiment is needed. For this purpose, multiple datasets reported by different groups for the sensitivity of $MoS_2$ gas sensors were collected and analyzed. We focused on sensors with monolayer $MoS_2$ sensing layers and the reported sensitivities for $NH_3$ as a gas acting as a donor and $NO_2$ behaving as an acceptor as two representative gases. Figure 5 shows a compilation of the results of our search and reveals a problem with the reported data. As can be seen, the sensitivities reported by different groups vary considerably even for the same gas at nominally identical concentrations, which complicates a direct comparison of the reported sensitivities of monolayer $MoS_2$ gas sensors [33–42].

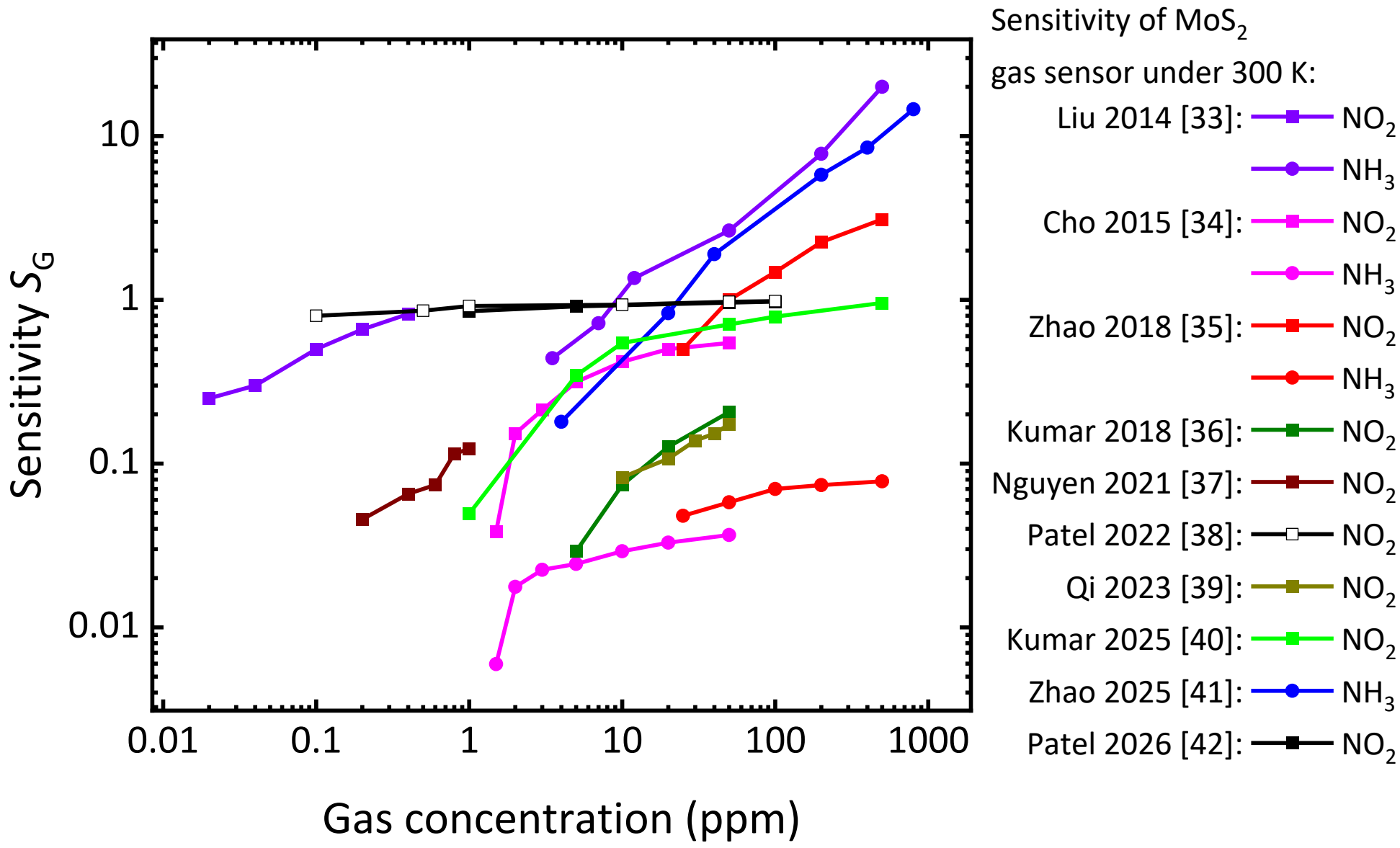

Figure 5. An overview of the sensitivity of experimental $MoS_2$ gas sensors [33–42].

For instance, when detecting $NO_2$ Cho et al. [34] reported a sensitivity of 50% at a $NO_2$ concentration of 20 ppm (parts per million), while Liu et al. [33] observed a similar sensitivity already for 0.1 ppm $NO_2$, corresponding to a concentration difference at the same sensitivity of approximately 200 times. Similarly, for a $NH_3$ concentration of 50 ppm, Cho et al. [34] measured an $S_G$ = 3.7%, whereas Liu et al. [33] reported an $S_G$ of 265%, differing by a factor of almost 72. These pronounced discrepancies significantly complicate a quantitative comparison between theory and experiment. The variations are most likely caused by differences in $MoS_2$ crystal quality, unintentional doping, and measurement setups. In addition, the sensor structures and measurement conditions are often unclear or incompletely described, which further complicates a meaningful quantitative comparison between theory and experiment.

An exception is the work by Liu et al. [33] which provides the sensor structure and measurement conditions needed for modeling and simulation, together with extensive experimental data on the sensor responses to $NH_3$ and $NO_2$. Therefore, the comparison between modeled and simulated sensitivities on the one hand and measured sensitivities on the other hand is conducted using the data reported in [33]. Table 3 summarizes the information relevant for sensor modeling and simulation.

| Temperature $T$ [K] | 300 | Thickness $t$ of $MoS_2$ [nm] | 0.65 |
|---|---|---|---|
| Doping $n_0$ [$cm^{-3}$] | $1.21 \times 10^{17}$ | Width $W$ of $MoS_2$ [µm] | 3.45 |
| Length $L$ of $MoS_2$ [µm] | 0.65 [33] | Thickness of $SiO_2$ $t_{SiO2}$ (nm) | 300 |

Table 3. Relevant information for modeling the sensitivity of the $MoS_2$ gas sensor [33]. Note: The background doping was not explicitly given in Ref. [33] but was inferred from the results reported therein.

As a first step, we use analytical modeling, i.e., approach (1), and the data sets for $E_{ads}$ and $\Delta Q$ from [50–53] shown above in Figs. 3 and 4, to calculate the sensitivity of the $MoS_2$ gas sensor from Ref. [33] for $NH_3$ adsorption. The aim of this study is to find out which of the available $E_{ads}$ and $\Delta Q$ data sets is most suitable to reproduce the experimental sensitivities. Table 4 shows the numerical values of $E_{ads}$ and $\Delta Q$ for $NH_3$ adsorption taken from [50–53].

| Reference | $E_{ads}$ (meV) | $\Delta Q$ |
|---|---|---|
| Late et al. [50] | −163 | 0.009 |
| Yue et al. [51] | −250 | 0.069 |
| Li et al. [52] | −407 | 0.0615 |
| Linghu et al. [53] | −185 | 0.031 |

Table 4. Adsorption parameters $E_{ads}$ and $\Delta Q$ for $NH_3$ adsorption on pristine $MoS_2$ reported in [50–53].

Figure 6 shows the modeled sensitivities of the $MoS_2$ gas sensor as a function of the $NH_3$ concentration in the sensor environment, together with the experimental sensitivities reported in [33] for the four $E_{ads} - \Delta Q$ data sets from Tab. 4. As to be expected due to the different values for $E_{ads}$ and $\Delta Q$ in Tab. 4, the modeled sensitivities show considerable differences.

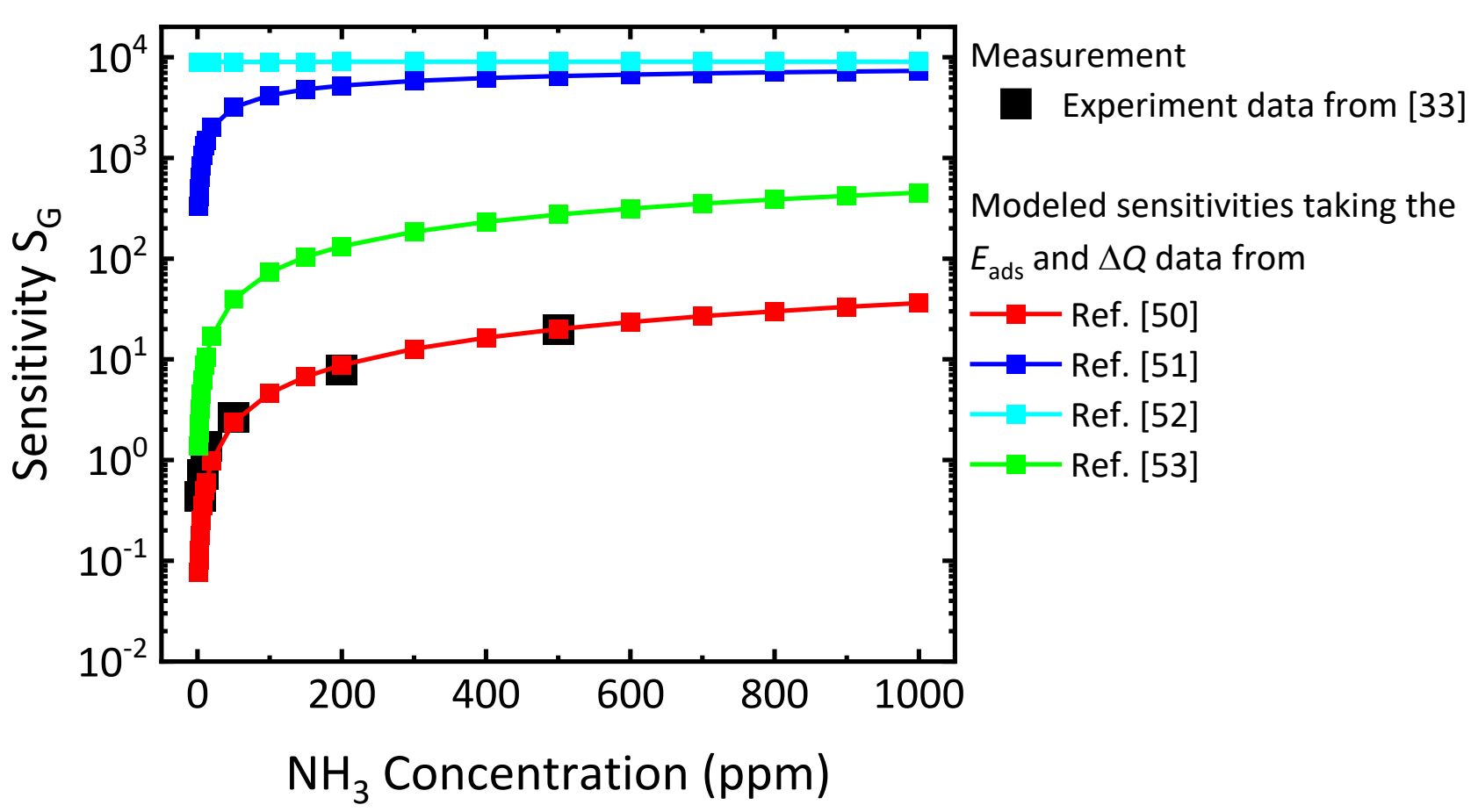

Figure 6. Modeled sensitivities of the $MoS_2$ gas sensor from [33] using the $E_{ads} - \Delta Q$ data sets from Tab. 4, together with the experimental sensitivities reported in [33], versus the $NH_3$ concentration in the sensor environment. For both the modeling and the experiment the applied voltages are $V_{GS}$ = 0 V and $V_{DS}$ = 1 V [50–53].

The measured sensitivities can be accurately reproduced using the $E_{ads}$ and $\Delta Q$ data from [50], whereas the corresponding data from [51–53] lead to a significant overestimation of the sensor sensitivity. This can be understood by the following consideration. As mentioned in Sec. 2.3, a higher absolute value of the adsorption energy means stronger adsorption. Let us take the $E_{ads}$ data from [51] as an example. Its value is -250 meV compared to the $E_{ads}$ of -163 meV reported in [50]. Thus, assuming the $E_{ads}$ from [51] a much higher density of adsorbed $NH_3$ molecules is predicted than in the case where the $E_{ads}$ from [50] is taken. Moreover, the charge transfer per adsorbed $NH_3$ molecule reported in [51] is larger than that from [50] (0.069 $q$ versus 0.009 $q$). The combined effect of the larger $|E_{ads}|$ and the larger $\Delta Q$ from [51] results in much larger $\Delta n$ and thus a much larger sensitivity, see eq. (9). A quantitative comparison of the sensitivities at 500 ppm $NH_3$ yields a measured sensitivity of 20, which is identical to the modeled sensitivity using the $E_{ads} - \Delta Q$ data from [50] compared to sensitivities of 275, 6500, and 9000 obtained for the $E_{ads} - \Delta Q$ data from [51–53].

In a second step, the sensitivities obtained by analytical modeling (approach 1) and device simulation (approach 2) are compared. We consider again the sensor structure from [33], $NH_3$ as the gas acting on the sensor, and the $E_{ads} - \Delta Q$ data for $NH_3$ from [50]. Figure 7 indicates that the sensitivities obtained by analytical modeling and those calculated from the currents simulated using ATLAS are in close agreement with the experimental data from [33].

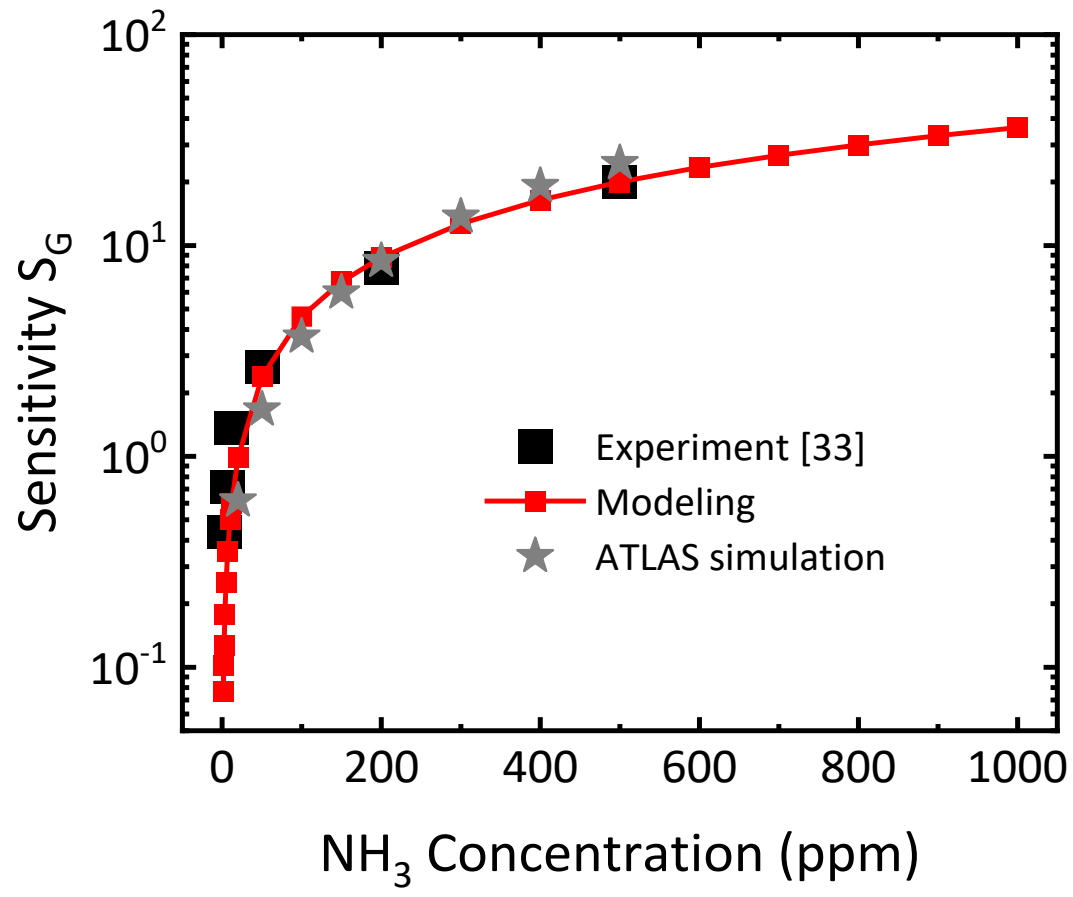

Figure 7. Comparison of the modeled, simulated, and experimental sensitivities of the $MoS_2$ gas sensors to $NH_3$ as a function of the $NH_3$ concentration for identical bias conditions ($V_{DS}$ = 1 V, $V_{GS}$ = 0 V).

Finally, the experimental sensitivity data of the monolayer $MoS_2$ gas sensor reported in [33] when exposed to $NO_2$, an acceptor-like gas, are compared to calculated sensitivities. Since, as shown in Fig. 7, modeling and simulation yield very similar results, we limit the considerations to analytical sensitivity modeling and refrain from computationally more expensive numerical simulations here.

In the experiments on $NO_2$ adsorption described in [33], between the back gate of the sensor and source a voltage $V_{GS}$ = 30 V was applied (for comparison: in the sensor measurements under $NH_3$ exposure $V_{GS}$ was zero), while all other parameters remained the same as in the $NH_3$ experiment. This gate bias alters the electron concentration in the $MoS_2$ channel by $\Delta n_0$ given by

$$\Delta n_0 = \frac{\Delta Q_{Ch}}{q \times V_{MoS_2}} = \frac{C_G \times \Delta V_{GS}}{q \times V_{MoS_2}} \tag{10}$$

where $\Delta Q_{Ch}$ is the change of the channel charge caused by a variation of the gate-source voltage, $V_{MoS2}$ is the volume of the $MoS_2$ channel (between source and drain) and $C_G$ is the (back-gate) capacitance, which is considered as a plate capacitor with the Si bulk as one plate, the $MoS_2$ channel as the second plate, and the $SiO_2$ in between as the dielectric. This leads to the expression

$$\Delta n_0 = \frac{\varepsilon_{SiO_2}}{q \times t_{SiO_2} \times t_{MoS_2}} \times \Delta V_{GS} \tag{11}$$

used to calculate the gate-induced charge modulation $\Delta n_0$. The overall electron concentration $n_0$ (30 V), i.e., the effective background doping at the gate-source voltage of 30 V used in the $NO_2$ experiment in [33], is obtained by adding the gate-induced charge modulation $\Delta n_0$ from (11) to the initial background doping concentration $n_0$ (0 V) used for the $NH_3$ modeling. The background doping for the $NH_3$ experiment, i.e., for zero $V_{GS}$, $n_0$ (0 V) is 1.21 × $10^{17}$ $cm^{-3}$, see Tab. 3, and an additional carrier concentration $\Delta n_0$ = 3.32 × $10^{19}$ $cm^{-3}$ is calculated for $\Delta V_{GS}$ = 30 V. The sum $n_0$ (0V) + $\Delta n_0$ represents the initial electron concentration and thus the background doping in the $MoS_2$ channel for the $NO_2$ experiment and amounts to ≈ 3.33 × $10^{19}$ $cm^{-3}$.

For acceptor-like gases such as $NO_2$, the calculation of the maximum adsorption density $\rho_{max}$ has to be reconsidered. As discussed above, acceptor-like gases extract electrons from the $MoS_2$ channel. When the gas concentration increases beyond a certain level, a situation may occur where all electrons are extracted from the channel and for higher gas concentrations the model would erroneously predict a transition of the conductivity of the $MoS_2$ channel from n-type to p-type. This limiting case occurs when the density of adsorbed gas molecules reaches a critical value $\rho_{crit}$ which for the case of $NO_2$ adsorption is given by

$$\rho_{\text{crit-NO}_2} = \frac{\left[n_0(0\,\text{V}) + \Delta n_0\right] \times t_{\text{MoS}_2}}{\left|\Delta Q_{\text{NO}_2}\right|} = 2.23 \times 10^{13}\,\text{cm}^{-2} \ll \rho_{\max} \tag{12}$$

Since this $\rho_{crit\text{-}NO2}$ value is much smaller than $\rho_{max}$ from eq. (4), when modeling the adsorption of $NO_2$ eq. (3) must be modified by replacing $\rho_{max}$ by $\rho_{crit\text{-}NO2}$.

After these preliminary considerations, we can now examine the adsorption of $NO_2$ and compare the experimental data from [33] with the results of our modeling. For the modeling, the $E_{ads}$ and $\Delta Q$ data from [50] ($E_{ads}$ = -285 meV and $\Delta Q$ = -0.097 × $q$, the minus sign indicates that $NO_2$ acts as an acceptor) are used, together with the gate-induced electron concentration variation $\Delta n_0$ = 3.32 × $10^{19}$ $cm^{-3}$ calculated above, $\rho_{max}$ = $\rho_{crit\text{-}NO2}$ = 2.23 × $10^{13}$ $cm^{-2}$ from Eq. (12), and a new fitting parameter $\beta$ = 0.8.

Figure 8 shows the modeled sensitivity and the experimental sensitivity data points from [33] and indicates a very good agreement for the sensor response across the entire $NO_2$ concentration range considered.

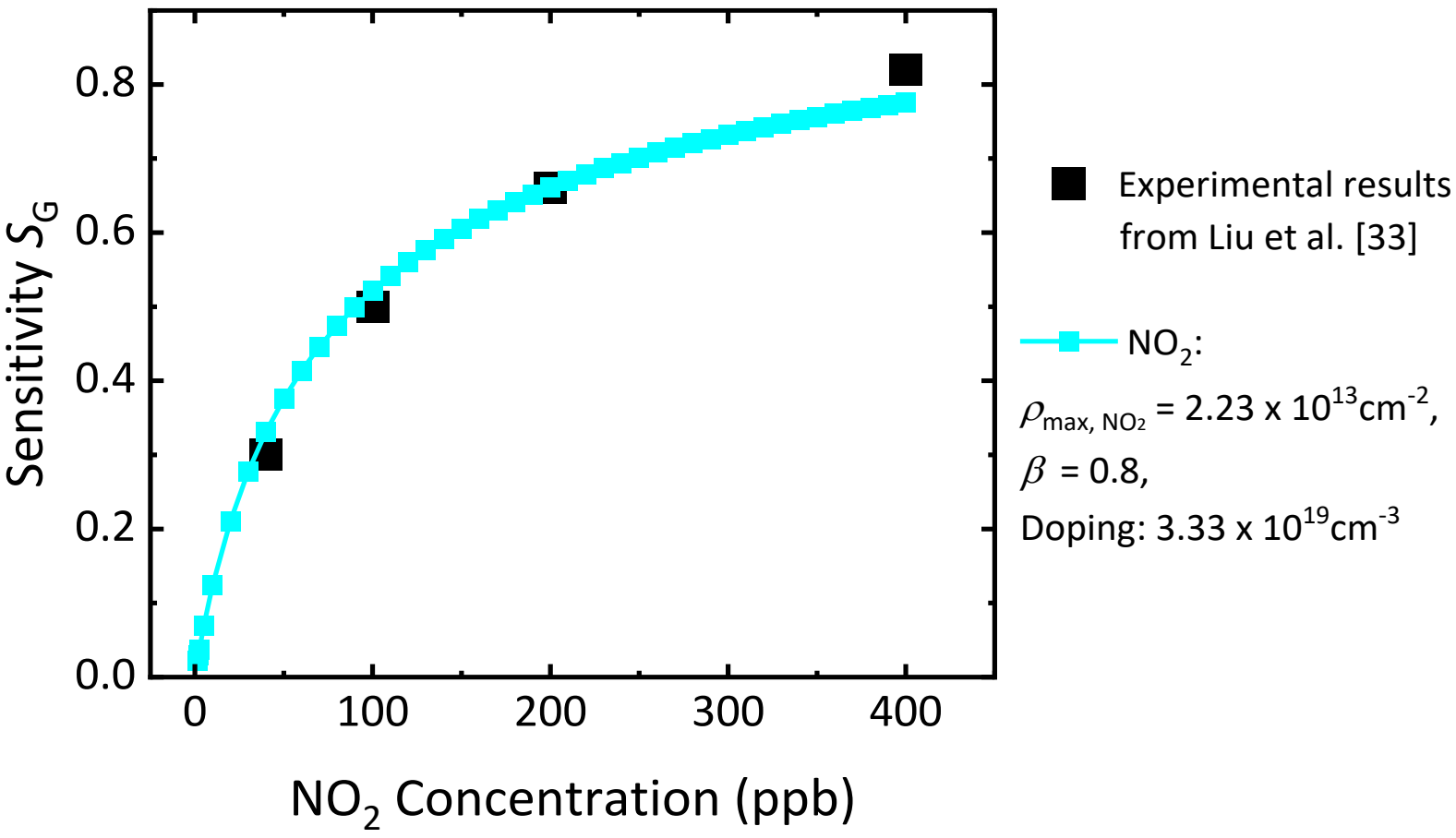

Figure 8. Modeled and measured sensitivity of the $MoS_2$ gas sensor from [33] for the adsorption of $NO_2$.

This result, together with the equally good agreement between modeling and experiment for $NH_3$ adsorption, demonstrates that the analytical model accurately reproduces the responses of $MoS_2$ gas sensors to both donor- and acceptor-type gases.

The comparison reveals that the relative deviation between modeling and experiment ranges from 0.15% to 10% across the examined $NO_2$ concentration range. At 40 ppb, the experimental and modeled sensitivities are $S_G$ = 0.30 and 0.33, respectively, corresponding to a deviation of 10 %. At 100 ppb, the sensitivities are $S_G$ = 0.50 and 0.525 with a deviation of 5 %, while at 200 ppb, $S_G$ = 0.66 and 0.661 with a deviation of 0.15 %. At the highest concentration of 400 ppb, the modeled sensitivity of $S_G$ = 0.78 closely matches the experimental value of $S_G$ = 0.82, yielding a deviation of 4.9 %.

### 4.2 Selectivity of $MoS_2$ gas sensors

A good gas sensor should not only show a high sensitivity but should also be able to distinguish between a specific target gas and other interfering gases in its environment. In other words, the sensor should respond selectively. A widely used definition of the selectivity $S_{sel}$ is [63]

$$S_{sel} = \frac{S_{G,\ target}}{S_{G,\ interfering}} \tag{13}$$

where $S_{G\text{-target}}$ is the sensor sensitivity to the target gas and $S_{G\text{-interferent}}$ is the sensitivity to a certain interfering gas.

| Gas species | $E_{ads}$ (meV) | $\Delta Q$ |
|---|---|---|
| $NO_2$ | −276 | −0.1 |
| CO | −128 | −0.02 |
| NO | −211 | −0.022 |
| $O_2$ | −116 | −0.041 |
| $H_2$ | −70 | −0.004 |
| $H_2O$ | −234 | −0.012 |
| $NH_3$ | −250 | 0.069 |

Table 5. The adsorption energy $E_{ads}$ and charge transfer $\Delta Q$ for different gas species reported by Yue et al. [51] (for $\Delta Q$, "−" denotes acceptor-like behavior, "+" denotes donor-like behavior).

In addition, since Late et al. [50] only provided $E_{ads}$ and $\Delta Q$ for $NH_3$ and $NO_2$, the analysis of gas selectivity requires a parameter set that consistently covers a wider range of gas species. For this purpose, the adsorption parameters reported by Yue et al. [51] are adopted, as they provide both $E_{ads}$ and $\Delta Q$ for multiple gases interacting with $MoS_2$, including $NO_2$, CO, NO, $O_2$, $H_2$, $H_2O$, $NH_3$, as summarized in Tab. 5. This enables a systematic and internally consistent evaluation of sensor selectivity under identical modeling assumptions.

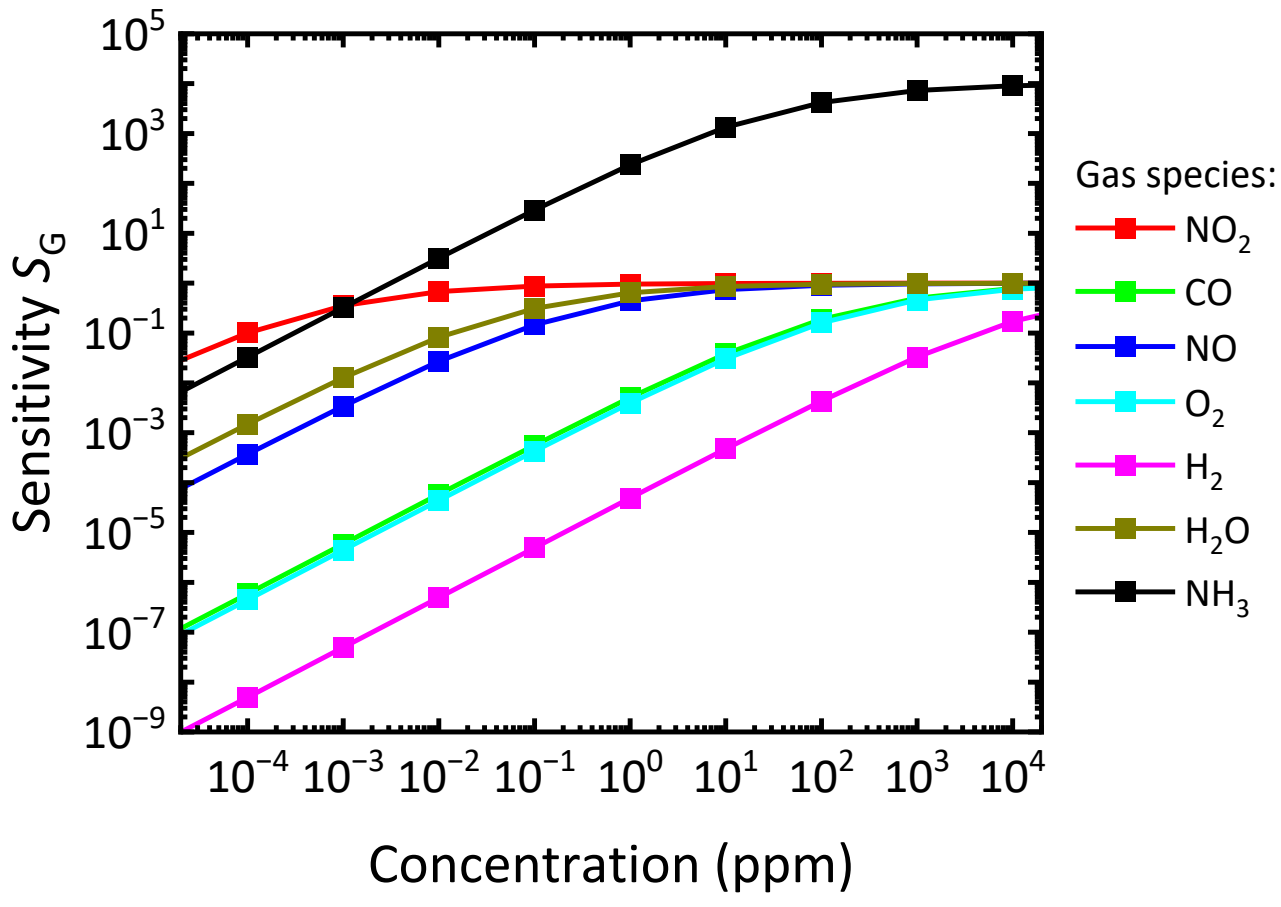

Figure 9. Sensitivity of a $MoS_2$ gas sensor for different adsorbed gas species.

The following selectivity analysis aims to capture relative trends under a consistent parameter set rather than to provide absolute quantitative predictions, since adsorption parameters reported in the literature may vary substantially. The calculated selectivity is based on the ratio of separately modeled single-gas responses; competitive adsorption and mixed-gas effects are not considered.

Based on the parameters summarized in Tab. 5, the calculated sensitivities $S_G$ for various gases are presented in Fig. 9. Except for $NH_3$, all analyzed gases exhibit acceptor-like behavior, where $S_G$ increases with concentration and gradually saturates near 100%. In contrast, $NH_3$, which behaves as a donor-like gas, produces a much stronger response, with $S_G$ exceeding 100% and reaching approximately $10^5$ at a concentration of $10^4$ ppm. Among the different gas species, $O_2$ (cyan) and CO (green) show nearly identical sensitivity values across the entire concentration range, for instance both reaching about 3 % at 10 ppm, thereby making discrimination based solely on sensitivity difficult. Meanwhile, $H_2$ exhibits significantly lower response, with $S_G < 20\%$ even at high concentrations of $10^4$ ppm.

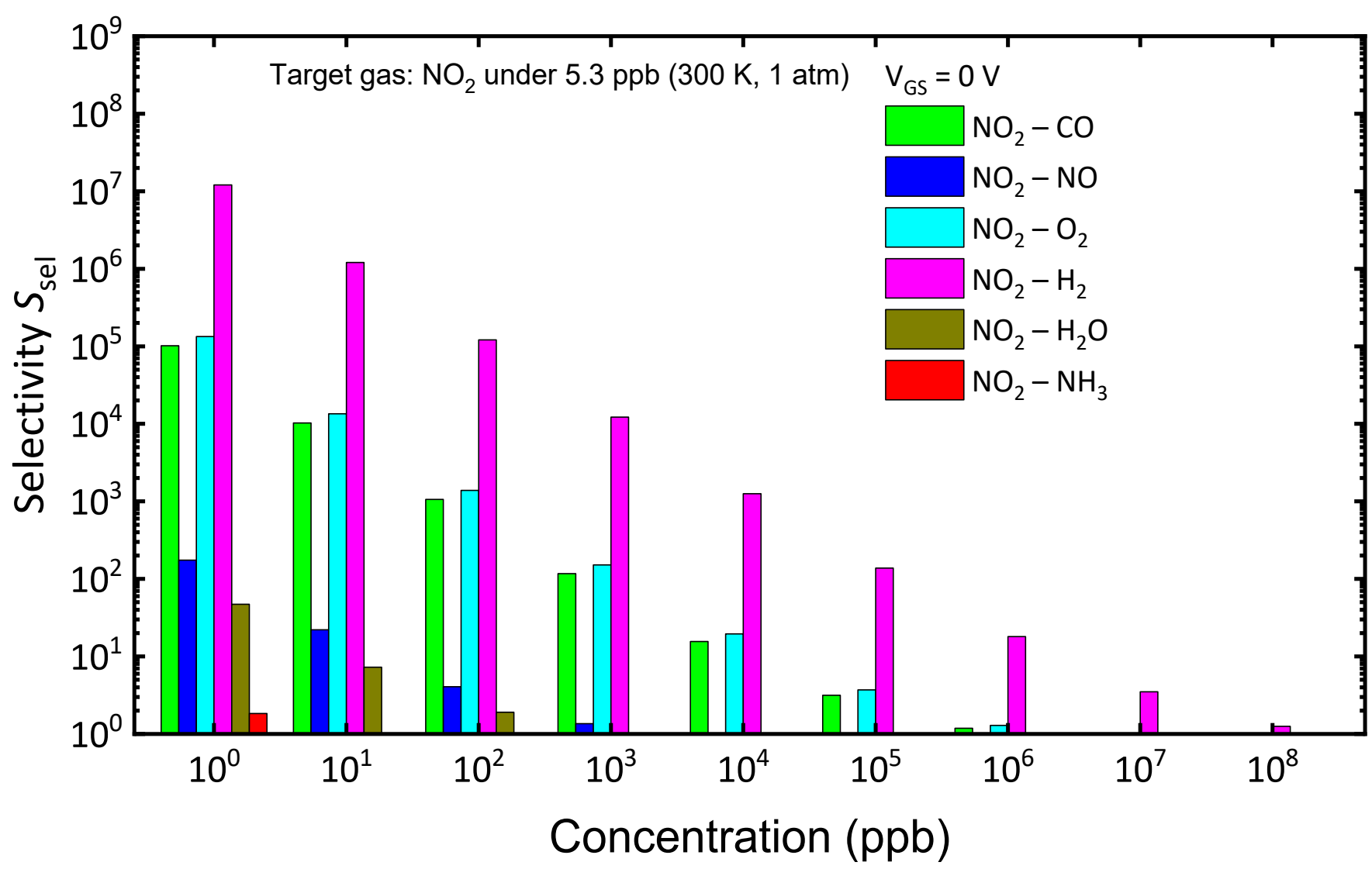

Figure 10. Ratio of $MoS_2$ gas sensor sensitivity at 5.3 ppb $NO_2$ relative to other gases, corresponding to the WHO annual mean air-quality guideline value (10 µg m$^{-3}$ ≈ 5.3 ppb) [2,51].

According to the updated 2021 WHO [2] guidelines for outdoor air pollution, the recommended annual mean $NO_2$ concentration is 10 µg m$^{-3}$, corresponding to approximately 5.3 ppb. Therefore, $NO_2$ was selected as the target gas for the selectivity analysis, while CO, NO, $O_2$, $H_2$, and $H_2O$ are considered as interfering gases. Figure 10 presents the selectivity $S_{sel}$ of $NO_2$ against CO, NO, $O_2$, $H_2$, and $H_2O$. At low concentrations, particularly at 1 ppb, the $MoS_2$ sensor exhibits strong discrimination, with $S_{sel}$ exceeding $10^7$ for $H_2$, remaining above $10^5$ for CO and $O_2$, and reaching 174 for NO and 47 for $H_2O$. These results indicate excellent selectivity toward $NO_2$ under low-concentration conditions. As the concentration of interfering gases increases, $S_{sel}$ decreases significantly, and at higher concentrations, the sensor's ability to distinguish $NO_2$ becomes limited. At $10^3$ ppb, it drops to about 117 for CO, 1.4 for NO, 151 for $O_2$, $10^4$ for $H_2$, and 0.9 for $H_2O$. In particular, $S_{sel}$ for NO and $H_2O$ approaches unity ($S_{sel}$ ≈ 1), suggesting that the sensor loses its ability to differentiate $NO_2$ from these gases. At higher concentrations, especially near $10^6$ ppb, $S_{sel}$ for CO, NO, $O_2$, and $H_2O$ all converge to approximately 1, while $H_2$ maintains a slightly higher $S_{sel}$ of about 17.9, which also decreases to near 1 at $10^8$ ppb.

These results suggest that as the concentration of interfering gases increases, the ability of the $MoS_2$ gas sensor to selectively detect $NO_2$ gradually weakens. In environments with high background gas concentrations, the selectivity of $MoS_2$-based gas sensors is therefore limited. Moreover, within the practically relevant low-concentration regime, the sensor maintains excellent selectivity toward $NO_2$, confirming its strong potential for air-quality monitoring applications.

## 5. Conclusion

A comprehensive Langmuir model has been developed to model the adsorption behavior and sensing characteristics of $MoS_2$-based gas sensors. By integrating adsorption parameters from Late et al. with the experimental configuration reported by Liu et al., the model accurately reproduces the sensitivity of both donor-type ($NH_3$) and acceptor-type ($NO_2$) gases across a wide concentration range. The analytical results exhibit excellent quantitative agreement with experimental data, confirming the validity of the Langmuir model in describing gas adsorption on two-dimensional materials. The model further enables quantitative evaluation of gas selectivity. The analysis demonstrates that $MoS_2$ exhibits strong selectivity toward $NO_2$ in the low-concentration regime relevant to WHO air-quality limits, while the selectivity gradually diminishes as the concentrations of the interfering gases increase and their modeled responses approach saturation. Overall, the proposed model provides a unified and physically consistent description of sensitivity and selectivity in TMDC-based gas sensors. This theoretical foundation can be extended to other 2D materials and gas species, offering a predictive approach for optimizing sensor design and improving detection performance in practical air-quality monitoring applications.